\documentclass[12pt]{iopart}

\usepackage{graphics}
\usepackage{latexsym}

\usepackage{iopams}

\begin{document}

\newtheorem{teo}{Theorem}
\newtheorem{lem}{Lemma}
\newtheorem{corol}{Corollary}
\newtheorem{defi}{Definition}
\newtheorem{esem}{Example}
\newtheorem{prop}{Proposition}
\newtheorem{rem}{Remark}

\newcommand\sh{{\rm sh}}
\newcommand\ch{{\rm ch}}
\newcommand\cth{{\rm cth}}
\newcommand{\beq}{\begin{equation}}
\newcommand{\eeq}{\end{equation}}
\newcommand{\R}{{\rm Re}}

\newcommand{\I}{{\rm Im}}
\newcommand{\Res}{{\rm Res}}

\title[]{Zeta function regularization
for a scalar field in a compact domain}

\author{G. Ortenzi\dag\ and M. Spreafico\ddag}

\address{\dag\ Dipartimento di Matematica ed Applicazioni, Universit\`a
Milano Bicocca, Italy, giolev@matapp.unimib.it}

\address{\ddag\ ICMC Universidade S\~{a}o Paulo, S\~{a}o Carlos, Brazil, mauros@icmc.usp.br}

\begin{abstract}
We express the zeta function associated to the Laplacian operator
on $S^1_r\times M$ in terms of the zeta function associated to the
Laplacian on $M$, where $M$ is a compact connected Riemannian
manifold. This gives formulas for the partition function of the
associated physical model at low and high temperature for any
compact domain $M$. Furthermore, we provide an exact formula for
the zeta function at any value of $r$ when $M$ is a
$D$-dimensional box or a $D$-dimensional torus; this allows a
rigorous calculation of the zeta invariants and the analysis of
the main thermodynamic functions associated to the physical models
at finite temperature.
\end{abstract}





\section{Introduction}
\label{s0} The zeta function regularization procedure is one of
the most elegant and rigorous methods to deal with path integrals
in quantum field theory. A great effort has been done in the last
years to produce exact calculations in different cases of interest
\cite{EORBZ} \cite{Eli} \cite{Eli10} \cite{EBK} \cite{BEGK} with
particular attention to exact calculation of the heat kernel
coefficients \cite{BEK}. Notice also some rigorous calculations
appeared in the recent mathematical literature: \cite{CQ}
\cite{Dow} \cite{Les} \cite{Spr1} \cite{Spr4} \cite{Ill}. The
purpose of this note is to show the result of a rigorous
application of the zeta function regularization technique. In
particular, we analyze the basic example considered by Hawking in
\cite{Haw}: a scalar field in a compact domain of the product
space time $S^1_{\beta/2\pi}\times M$ at temperature
$T=\frac{1}{\beta}$, and provide a complete treatment of it. Our
main motivation is that we can perform all the calculations in a
rigorous way and determinate all the quantities of interest
without introducing any approximation. Our main result is an
effective formula expressing the partition function at finite
temperature in terms of the geometry of the spatial background.
This allows on one side to describe the behavior of the physical
model at low and high temperature for any spatial domain (cfr
with \cite{CKVZ}, where a similar analysis was performed for
$M=H^3/\Gamma$); on the other side, to provide explicit formulas
that describe the model for some particular geometries, namely
the $D$-dimensional box and the $D$-dimensional torus, at any
value of the temperature. Our technique is likely to be
generalized to other situations, and there are works in progress
in various directions.

\noindent The partition function for a scalar field $\phi$ with action
$I[\phi]=-\int\phi(x) A\phi(x)d(vol(x))$ in the Euclidean space time
and where $A$ is a real elliptic self adjoint second order
differential operator, can be formally described by the Feynman path
integral $ Z=\int \mathcal{D}\phi \e^{\rho I[\phi]}=[{\rm det}(\rho
A)]^{-\frac{1}{2}}$ where $\rho$ is some renormalization constant with
dimension of mass or inverse length introduced by hand in order to
obtain a physically consistent model \cite{Haw}\footnote{Beside in
this work we will not treat renormalization problems,  we will give
effective formulas for $Z$ for any $\rho$.}. With periodic boundary
condition in the imaginary time with period $\beta$, the background
geometry of the flat space time is described by the product space
$S^1_{\beta/2\pi}\times M$, with product metric $1\oplus g$, where
$(M,g)$ is a complete Riemannian manifold of dimension $D$; in
particular, we assume $M$ to be compact connected. Following Hawking
\cite{Haw}, the path integral can be given a rigorous interpretation
in terms of some zeta invariants associated to the underlying
geometry. We introduce the zeta function associated to the operator
$A$, $\zeta(s;A)=\sum_{\lambda \in{\rm Sp_0}A}\lambda^{-s}$,  (where
${\rm Sp}_0A$ denotes the non vanishing part of the spectrum of $A$);
then, defining the regularized determinant of $A$ \cite{RS} \cite{ABP}
by $\log{\rm det} A=-\left. \frac{d}{ds}\zeta(s;A)\right|_{s=0}$,  we
get
$\log Z=\frac{1}{2}\zeta'(0;A)
-\frac{1}{2}\log \rho \zeta(0;A)$. Under the identification
$T=\frac{1}{\beta}$, the partition function for the quantum
theoretical model corresponds to the one for a canonical ensemble at
temperature $T$. We will assume this point of view, and we will work
out the partition function $Z_T$ for a statistical system calculating
the zeta invariants of the underlying geometry. This will allow us to
introduce and analyze other interesting thermodynamic functions and to
get useful information on the system described at finite temperature
$T$. The main feature of this approach is that we can describe
completely the partition function for any theory in the product space
time at any  value of the temperature, in terms of a zeta function at
null temperature (that we will call geometric zeta function) depending
only on the geometry of the space, namely of the background physical
domain. To get this result, after introducing the zeta function
associated to the model in the proper Euclidean setting, we use the
particular form of the spectrum to decompose the zeta function. We
easy get a function of the temperature $T$, smooth for all positive
$T$, but the hard point is to get the right analytic extension at
$s=0$, in order to obtain the partition function. In particular, the
analytic continuation is not uniform in $T$ for small $T$, and hence
the calculation of $\zeta'(0)$ must be performed at positive $T$. We
are able to get this result, and we can do it uniformly in $T$ for $T$
in any closed interval of the real positive axis (Proposition \ref{p2}
of Section \ref{s1}). Moreover, the result is effective, and we can
deduce from it both the behaviors for low and high temperature.

\noindent Studying the models at low temperature, we show that periodic
boundary conditions on the spatial domain give an anomalous behavior
for the zero rest mass scalar field that can be corrected by adding a
non zero mass term, while studying the behavior of the pressure of the
radiation as a function of the volume at different fixed temperatures,
we prove the existence of a minimal non vanishing value of the volume
below which the force becomes attractive, as expected because of the
Casimir effect \cite{Bor} \cite{Mil}. More precisely, the expansions
at low temperature of the thermodynamic functions show the following
effect of the boundary conditions on the physics of the model: the
presence of a spatial zero mode produces an anomalous behavior of the
entropy and the specific heat at low temperature. In particular, on a
closed (compact, connected, with no boundary) geometry, we always have
a zero mode for the Laplacian, due to the constant eigenfunction, and
this produces a $\log T$ term in the logarithm of the partition
function, and hence a logarithmic divergence in the entropy and a non
zero specific heat at low $T$. Such a term disappears when a boundary
is present and opportune boundary conditions are assumed. Furthermore,
the presence of a non zero mass term cancels the logarithmic term and
recasts a correct behavior for small $T$ on both the domains,
independently from the boundary conditions. This is in agreement with
the appearance of a logarithmic divergence (infrared divergence) for a
massless field classically (see for example \cite{Ram}). Particularly
meaningful in this context is the case $D=1$, whose associated zeta
function appears as the zeta function for the Laplacian with a constant
potential on a cylinder or on a torus. Its $s$-expansion near $s=0$ is
well known since Eisenstein and Kronecker \cite{Wei}, and relates to
the theory of modular functions and $L$-series. The coefficient of the
linear term can be expressed using the Dedekind eta function, and
using the modular property of the last we can easily relate the
behavior at low and high temperature of the partition function of the
physical model. It is clear why the presence of a mass term cancels the
singular behavior at low $T$: in fact, such a non homogeneous term
breaks the modular property of the eta function. This will be
discussed at the end of part \ref{s33}. The $D$ dimensional version of
the eta function appearing in the corollary of Proposition \ref{p2},
has not been studied yet. It does not exhibit any periodicity, but
possible modular properties should be very important. A deeper
analysis of such function should be very interesting both under a
theoretical and applicative point of view.

\noindent The study of the thermodynamic function at finite
temperature allows to analyze other interesting phenomena. In
particular, we investigate the pressure of the radiation at finite
temperature as a function of the volume, and we show the existence,
for any fixed temperature, of a critical volume $V_0$ where the
pressure of the radiation changes sign, becoming attractive. The
analysis performed in part \ref{s35} of Section \ref{s3} shows that
below $V_0$ the pressure is attractive and decreases like
$V^{-\frac{D+1}{D}}$, as expected by the Casimir effect \cite{Bor}
\cite{Mil}. We also show that, up to renormalization, the critical
volume depends on the temperature by the law $T_0 V_0^\frac{1}{D}={\rm
const}$, and the boundary does not affect this effect.

\noindent The work is organized as follows. In Section \ref{s1} we
briefly recall some basic information about the geometry of the
Laplace operator on compact manifolds. This will lead us to define
the zeta function that describes the physical models we want to
study and to state our main result in Proposition \ref{p2} and its
corollary. In Section \ref{s2}, we prove asymptotic expansions at
high and low temperature. In Section \ref{s3}, we study the zeta
function introduced in Section \ref{s1} for some particular
geometries at finite temperature, and we provide formulas for the
main thermodynamic functions.

\section{The zeta regularized partition function}
\label{s1}

We recall some known facts about the geometry of the Laplace
operator on compact manifolds. Let $(M,\partial M,g)$ be a compact
connected Riemannian manifold of dimension $D$ with (possibly
empty) boundary $\partial M$. Let $(\Delta_M,BC)$ be the Laplace
operator built using the metric $g$ and acting on the opportune
space of functions with boundary conditions $BC$ on the boundary
of $M$. If the boundary of $M$ is not empty, Dirichlet or Neumann
boundary conditions can be chosen. Let ${\rm Sp}(-\Delta_M, BC)$
denotes the spectrum of the associated boundary value problem.
Then, ${\rm Sp}(-\Delta_M, BC)$ is a discrete set of non negative
(real) numbers: $\{\lambda_k\}_{k\in K}$, $K\subseteq \mathbb{N}$,
and ${\rm ker}(-\Delta_M)$ is a finite set of rank $\mathcal{K}$.
Let ${\rm Sp}_0$ denotes the set of the positive eigenvalues and
$K_0$ the set of their indices. If $\partial M$ is empty, then
$\mathcal{K}=\#ker(-\Delta_M)=1$, the constant function. In this
situation, we have the Weyl formula, that gives the behavior of
the large eigenvalues, $\lambda_k\sim\frac{4\pi^2}{({\rm vol}M
{\rm vol} B_D)^\frac{2}{D}} k^\frac{2}{D}$, where $B_D$ is the
unit ball in $\mathbb{R}^D$, and ${\rm
vol}B_D=\frac{\pi^\frac{D}{2}}{\Gamma\left(\frac{D}{2}+1\right)}$,
\cite{LY} \cite{Mel} \cite{RSiv} \cite{Hor}, and the heat kernel
expansion $ \sum_{k\in K} \e^{-\lambda_k t} =
t^{-\frac{D}{2}}\sum_{j=0}^\infty e_j t^\frac{j}{2}$, when $t\to
0^+$ \cite{Gil}. The formulas above hold also if a regular
potential term is added to the Laplacian and $e_0=\frac{{\rm
vol}M}{(4\pi)^\frac{D}{2}}$ \cite{Gil}. Before introducing the
zeta function, we need the following technical lemma, that can be
easily deduce from the Young's inequality.
\begin{lem} For all real positive $a$ and $b$, integer positive $n$ and $k$,
$ n^a+k^b>(nk)^\frac{ab}{a+b}$. \label{l0}
\end{lem}
We are now able to introduce the associated zeta function. We want
to do this in slightly more general setting, namely we assume a
possible mass term, and we consider the operator $-\Delta_M+q$,
with real $q\geq 0$. The kernel of this operator depends on $q$,
but its rank is discontinuous at $q=0$. We will see in the
following, that all the results we are able to prove, are true
uniformly in $q$, for all $q$ in some fixed closed interval of the
positive real axis. This suggests to deal independently with the
zero mass case. On the other side, to simplify notation and avoid
to give always two formulas, it is preferable to choose the
following alternative approach. Let's introduce the function $
\mathcal{K}_q=\left\{\begin{array}{ll}0&{\rm if}~q\not=0\\
\mathcal{K}&{\rm if}~q=0,\end{array}\right. $, and assume $q$ to
be fixed. Then, we can write a unique formula for both the cases,
$q=0$ and $q\not=0$, that is clearly non smooth in $q$. Also, we
state now once for ever that all the following formulas are smooth
and uniform in $q$ for $q$ in any fixed closed interval of the
positive real axis.

\noindent We first introduce the zeta function at null temperature,
\[
\zeta(s;0,q)
=\zeta(s;-\Delta_M+q,BC)
={\sum_{k\in K}}' (\lambda_k+q)^{-s},
\]
for $\R(s)>\frac{D}{2}$, where the notation means that the (possible)
zero terms must be omitted in the sum; we will call this function the
geometric zeta function associated to the model. Next, consider the
product manifold $S^1\times M$, with the product metric $d^2t\oplus
g$, Laplacian $\Delta=\frac{\partial^2}{\partial^2 t}+\Delta_M$, and
periodic boundary conditions $u(0)=u(\beta)$ on $S^1$ parameterized by
$t\in [0,\beta]$. Then, ${\rm Sp}(-\Delta, BC)=\{ (2\pi T
n)^2+\lambda_k\}_{n\in \mathbb{Z}, k\in K}$, and the associated zeta
function is\footnote{Notice that the same consideration introduced
previously about the variable $q$ apply for the variable $T$, but
recalling the discussion outlined in the introduction, we will assume
positive definite temperature and will work out the $T=0$ limit once
the right analytic continuation has been achieved.}
\[
\zeta(s;2\pi T,q) ={\sum}'_{(n,k)\in \mathbb{Z}\times K} [(2\pi
T n)^2+\lambda_k+q]^{-s},
\]
for $\R(s)>\frac{D+1}{2}$ by Lemma \ref{l0}. The zeta regularized
partition functions for a (possibly) massive scalar field whose
underlying geometry is $(M,\partial M,g)$ is
$$
\log Z(T,q) =-\frac{1}{2}\log \det
(\rho(-\Delta+q),BC)=\frac{1}{2}\zeta'(0;2\pi T,q)
-\frac{1}{2}\log\rho \zeta(0;2\pi T,q).
$$
As stated in the introduction, our approach is to study the zeta
functions as mathematical objects, providing all the zeta
invariants, and hence to write down formulas for the partition
function of the physical models and to calculate the thermodynamic
functions. In the most general setting of this section, all basic
information about the zeta function can be obtained using
classical methods (see \cite{Gil} or \cite{Ros}). We summarize
them in Propositions \ref{p1a} and \ref{p1b}. We also provide the
fundamental analytic representation for the zeta function,
Proposition \ref{p2}, that will be the starting point for all the
proofs in the following. From now on, we will use the variable
$y=2\pi T$ to simplify notation.
\begin{prop} The function  $\zeta(s;0,q)$
has an analytic continuation to the whole complex $s$-plane up to a
set of simple poles at the value of $s=\frac{D-l}{2}$, for
$l=0,1,2,\dots$, that are not non-positive integers,  with residua:
${\rm Res}_1\left(\zeta(s;0,q),s=\frac{D-l}{2}\right)
=\frac{1}{\Gamma\left(\frac{D-l}{2}\right)} \sum_{j,k\geq 0, j+2k=l}
\frac{(-1)^k}{k!}e_jq^k$; at zero and negative integers,
$-m=0,-1,-2,-3,\dots$: $\zeta(-m;0,q) =(-1)^m m!\sum_{j,k\geq 0,
j+2k=2m+D} \frac{(-1)^k}{k!}
e_jq^k-\frac{\mathcal{K}_q}{\Gamma(1-m)}$. \label{p1a}
\end{prop}
\begin{prop} The function  $\zeta(s;y,q)$
has an analytic continuation to the whole complex $s$-plane,
uniformly in $y$ for $y$ in any closed interval of the positive
real axis, up to a set of simple poles at the value of
$s=\frac{D+1-l}{2}$, for $l=0,1,2,\dots$, that are not
non-positive integers,  with residua\footnote{A further term
$\frac{2\mathcal{K}_q}{y}$ appears at $s=\frac{1}{2}$.}: ${\rm
Res}_1\left(\zeta(s;y,q),s=\frac{D+1-l}{2}\right)
=\frac{\sqrt{\pi}}{\Gamma\left(\frac{D+1-l}{2}\right)}\frac{1}{y}
\sum_{j,k\geq 0, j+2k=l} \frac{(-1)^k}{k!}e_jq^k$; $s=0$ is a
regular point and $\zeta(0;y,q)
=\frac{\sqrt{\pi}}{y}\sum_{j,k\geq 0, j+2k=D+1} \frac{(-1)^k}{k!}
e_jq^k-\mathcal{K}_q$. \label{p1b}
\end{prop}
Notice in particular the homogeneous case $
\zeta(0;0,0)=e_D-\mathcal{K}$,
$\zeta(0;y,0)=\frac{\sqrt{\pi}}{y}e_{D+1}-\mathcal{K}$.
\begin{prop} For all $y>0$, fixed $q\geq 0$ and uniformly in $s$ near $s=0$,
\[
\zeta(s;y,q)=\frac{\sqrt{\pi}}{\Gamma(s)}y^{-1}
\Gamma\left(s-\frac{1}{2}\right)\zeta\left(s-\frac{1}{2};0,q\right)
+2\mathcal{K}_q y^{-2s}\zeta_R(2s)+
\]
\[
+\frac{4\pi^s}{\Gamma(s)} y^{-s-\frac{1}{2}} \sum_{n=1}^{\infty}
{\sum_{k\in K} }'
\left(\frac{n}{\sqrt{\lambda_k+q}}\right)^{s-\frac{1}{2}}
K_{s-\frac{1}{2}}\left(\frac{2\pi
n}{y}\sqrt{\lambda_k+q}\right).
\]
\label{p2}
\end{prop}
\noindent\underline{Proof} Let $a_k=\lambda_k+q$.
First, we isolate the (possible) vanishing terms:
\[
\zeta(s;y,q)=\sum_{n=-\infty}^{+\infty} {\sum_{k\in K}}'[(y
n)^2+a_k]^{-s} +2\mathcal{K}y^{-2s}\zeta_R(2s);
\]
the second term needs no
further comments, for what concerns the first, we proceed as follows.
To start, assume $y>0$ to be fixed. In such a case, we provide a formula that can be
analytically continued in $s$ near $s=0$, and we prove that such extension is
uniform in $y$ for all $y\geq 0$.
We apply first the Mellin transform,
\[
\sum_{n=-\infty}^{+\infty} {\sum_{k\in K}}'[(y n)^2+a_k]^{-s}
=\frac{1}{\Gamma(s)}\int_0^\infty t^{s-1}
\sum_{n=-\infty}^{+\infty}\e^{-y^2n^2t} {\sum_{k\in K}}'
\e^{-a_kt}dt,
\]
and hence the multi dimensional Poisson summation formula, we get
\[
\frac{\sqrt{\pi}}{y\Gamma(s)}\int_0^\infty t^{s-\frac{1}{2}-1}
{\sum_{k\in K}}' \e^{-a_kt}dt
+\frac{2\sqrt{\pi}}{y\Gamma(s)}\int_0^\infty t^{s-\frac{1}{2}-1}
\sum_{n=1}^{\infty}\e^{-\frac{\pi^2 n^2}{y^2t}}
{\sum_{k\in K}}' \e^{-a_kt}dt=
\]
the integrals are known and give (where $K_\nu(z)$ is the Bessel
function)
\[
\hskip -2.55truecm=\frac{\sqrt{\pi}}{y\Gamma(s)}
\Gamma\left(s-\frac{1}{2}\right)\zeta\left(s-\frac{1}{2};0,q\right)
+\frac{4\pi^s y^{-2-\frac{1}{2}}}{\Gamma(s)}
\sum_{n=1}^{\infty}{\sum_{k\in K}}'
\left(\frac{n}{\sqrt{a_k}}\right)^{s-\frac{1}{2}}
K_{s-\frac{1}{2}}(2\pi n y^{-1}\sqrt{a_k}).
\]
The second term is a integral function of $s$, and it is also
easy to see (using classical estimates for the Bessel functions)
that convergence of the series is uniform in $y$ for bounded
$y\geq 0$. This means that the second term can be analytically
extended for all $s$ smoothly in $y\geq 0$. The first term can
have poles, but dependence on $y$ and $s$ are clearly distinct.
Using the results in Proposition \ref{p1b}, and the known
expansion for the Gamma function, we find out that there exists a
closed neighborhood of $s=0$ where there are no poles
independently from $y$, for all $y\geq 0$. This gives the thesis.
$\Box$

\noindent In the following result is the main decomposition of the
partition function of the physical model at any given temperature in
term of the geometric zeta function.
\begin{corol}
$$
\hskip -2cm \log Z(T,q) =\#{\rm ker}(-\Delta_M+q)\log T
-\log{\prod_{k\in K} }'
\left(1-\e^{-\frac{2\pi}{y}\sqrt{\lambda_k+q}}\right)+
$$
$$
\hskip 0.5cm -\frac{1}{2T}{\rm
Res}_0\left(\zeta(s;0,q),s=-\frac{1}{2}\right) -\frac{1}{T}(1-\log
2){\rm Res}_1\left(\zeta(s;0,q),s=-\frac{1}{2}\right)+
$$
\[
\hskip 2cm -\frac{1}{2}\log\rho\zeta(0;2\pi T,q),
\]
smoothly in $T>0$ and for all fixed $q\geq 0$.
\end{corol}
\noindent\underline{Proof} From Proposition \ref{p2}, we just
have to work out the first term. Recall that $
\frac{1}{\Gamma(s)}=s+\gamma s^2+O(s^3) $, and
$\Gamma\left(s-\frac{1}{2}\right)
=-2\sqrt{\pi}-2\sqrt{\pi}[2(1-\log 2)-\gamma]s+O(s^2)$,  and use
Proposition \ref{p1a} to write
\[
\zeta\left(s-\frac{1}{2};0,q\right)=\frac{c_{-1}}{s}+c_0+c_1s +O(s^2)
=\frac{R_1}{s}+R_0+c_1s +O(s^2),
\]
when $s\to 0$, where $R_i={\rm
Res}_i\left(\zeta(s;0,q),s=-\frac{1}{2}\right)$. Then, $
\frac{1}{\Gamma(s)}\Gamma\left(s-\frac{1}{2}\right)
\zeta\left(s-\frac{1}{2};0,q\right)
=-2\sqrt{\pi}R_1-2\sqrt{\pi}[R_0+2(1-\log 2)R_1]s+O(s^2)$, and
the corollary follows. $\Box$

\section{Low and high temperature expansions}
\label{s2}

We state our results in the most general setting, using the zeta
function introduced in the previous section. The expansions for
the partition function at low and high temperature can be
immediately obtained as corollaries (For an overview on the
physical literature on the thermodynamic on the zero mode and the
high and low temperature expansions the interested reader can
have a look at \cite{ET} and to the references there).

\begin{prop} For $y\to 0^+$, with fixed $q\geq 0$ and uniformly in $s$ near
$s=0$,
\[
\hskip -1.5truecm\zeta(s;y,q)
=\frac{\sqrt{\pi}y^{-1}}{\Gamma(s)}\Gamma\left(s-\frac{1}{2}\right)\zeta\left(s-\frac{1}{2};0,q\right)
+2\mathcal{K}_q y^{-2s}\zeta_R(2s) +O\left(\e^{-\frac{1}{y}}\right).
\]
\label{p3}
\end{prop}
\noindent\underline{Proof} This follows immediately from Proposition \ref{p2}.
$\Box$
\begin{corol} For small $T$, and all fixed $q\geq 0$,
\[
\hskip -1.5truecm \log Z(T,q) =\#{\rm ker}(-\Delta_M+q) \log T
-\frac{1}{2T}{\rm
Res}_0\left(\zeta(s;-\Delta_M,BC),-\frac{1}{2}\right)+
\]
\[
-(1-\log 2) {\rm
Res}_1\left(\zeta(s;-\Delta_M,BC),-\frac{1}{2}\right)\frac{1}{T}+
\]
\[
-\frac{1}{2}log\rho\left(\frac{1}{\sqrt{\pi}T}\sum_{j,k\geq 0,
j+2k=D+1} \frac{(-1)^k}{k!} e_j q^k-\mathcal{K}_q \right)
+O\left(\e^{-\frac{1}{T}}\right).
\]
\end{corol}
\begin{prop} For $y\to +\infty$, with fixed $q\geq 0$, and uniformly in $s$ near $s=0$,
\[
\hskip -2.5truecm\zeta(s;y,q)= \zeta(s;0,q)
+\frac{2e_0\pi^{-\frac{1}{2}+2s-D}}{\Gamma(s)}
\Gamma\left(\frac{D+1}{2}-s\right)\zeta_R(D+1-2s)y^{D-2s} +o(y^{D-2s}).
\]
\label{p4}
\end{prop}
\noindent\underline{Proof} Let $a_k=\lambda_k+q$.
We first isolate the (possible) vanishing terms as
follows:
\[
\zeta(s;y,q) =2\sum_{n=1}^{\infty} \sum_{k\in K}[(y n)^2+a_k]^{-s}
+\zeta(s;0,q);
\]
let $\sigma=\frac{1}{y^2}$, then:
$\zeta^{(1)}(s;y,q)=y^{-2s}2\sum_{n=1}^{\infty} \sum_{k\in K}
[n^2+\sigma a_k]^{-s} =y^{-2s}\zeta_\sigma(s;q)$, and we study
$\zeta_\sigma(s;q)$. We start assuming $\R(s)\geq
s_1>\frac{D+1}{2}$, $\sigma\in [\sigma_1,1]$. Since convergence is
absolute we can exchange summation indices as desired. Applying the
Mellin transform
\[
\zeta^{(1)}_\sigma(s;q) =\frac{1}{\Gamma(s)}\int_0^\infty
t^{s-1} 2\sum_{n=1}^\infty \e^{-n^2 t} \sum_{k\in K} \e^{-\sigma
t a_k} dt;
\]
splitting the integral at $t=\frac{1}{\sigma}$, in the first integral,
$\sigma t\leq 1$, and hence we can use the heat kernel expansion for
$-\Delta_M$, while the second integral gives a regular function of $s$
for all $s$. For the first term, we get (recall $\alpha=\frac{D}{2}$)
\[
\frac{\sigma^{-\alpha}}{\Gamma(s)}\sum_{j=0}^\infty e_j\sigma^\frac{j}{2}
2\sum_{n=1}^\infty
\int_0^{\frac{1}{\sigma}} t^{s+\frac{j}{2}-\alpha-1} \e^{-n^2 t}dt;
\]
notice that we can not use the Poisson formula here, since we are
interested in the small $\sigma$ expansion. Now, for each $j$,
\[
\hskip -2.2truecm \sum_{n=1}^\infty \int_0^{\frac{1}{\sigma}}
t^{s+\frac{j}{2}-\alpha-1} \e^{-n^2 t}dt
=\Gamma\left(s-\alpha+\frac{j}{2}\right)\zeta_R(2s-2\alpha+j)
-\sum_{n=1}^\infty \int_{\frac{1}{\sigma}}^\infty
t^{s+\frac{j}{2}-\alpha-1} \e^{-n^2 t}dt.
\]
The first term can just have simple poles at
$\alpha-\frac{j}{2}$ and $\alpha-\frac{1-j}{2}$. In fact the Gamma
function can have a pole only at non positive integer argument, but
for even negative integers the zeta function vanishes. Thus, for each
$j$ and all $s\not =\alpha-\frac{j}{2}, \alpha-\frac{1-j}{2}$,
\[
\lim_{\sigma\to 0^+}2\sum_{n=1}^\infty
\int_0^{\frac{1}{\sigma}} t^{s+\frac{j}{2}-\alpha-1} \e^{-n^2 t}dt
=2\Gamma\left(s-\alpha+\frac{j}{2}\right)\zeta_R(2s-2\alpha+j),
\]
where the analytic extension of the above function of $s$ is intended for
$\R(s)<\alpha+\frac{j}{2}$.
For small $\sigma$, we can write
\[
\hskip -2.6truecm \frac{1}{\Gamma(s)}\int_0^{\frac{1}{\sigma}} t^{s-1}
\sum_{n=1}^\infty \e^{-n^2 t} \sum_{k=0}^\infty \e^{-\sigma t a_k(x)}
dt =\sum_{j=0}^\infty e_j\sigma^{\frac{j}{2}-\alpha}
\frac{\Gamma\left(s-\alpha+\frac{j}{2}\right)\zeta_R(2s-2\alpha+j)}{\Gamma(s)}+o(1)
\]
and this is a regular function of $s$ for $s$ closed to $0$,
independently from $\sigma$, and hence gives the desired result. The
second term with $x=\sigma t$, and using the Poisson formula
becomes
\[
\sigma^{-s}\int_1^\infty x^{s-1} 2\sum_{n=1}^\infty \e^{-\frac{n^2
x}{\sigma}} \sum_{k=0}^\infty \e^{-x a_k(x)} dx= \sigma^{-s}
O(\e^{-\frac{1}{\sigma}}),
\]
for all $s$, since the second factor is a regular function of $s$ for
all $s$. Collecting and recasting the correct functions we get the
thesis, where we put in evidence the leading term and use the
Riemann's functional equation to rewrite its coefficient. $\Box$
\begin{corol} For large $T$,
\[
\log Z(T,q)=
\frac{{\rm vol}M}{\pi^{\frac{D+1}{2}}}
\Gamma\left(\frac{D+1}{2}\right)\zeta_R(D+1)T^D+o(T^D).
\]
\end{corol}

\section{Finite temperature results for some particular geometries}
\label{s3} We turn now our attention to the case of a scalar field of
rest mass $q$ in two fixed geometries: a cubic box $\mathcal{B}_D$
of edge $l$ and a $D$-dimensional torus
$\mathcal{T}_D=S_{l/\pi}^1\times\dots \times S_{l/\pi}^1$, at any
fixed value of the temperature $T$. The operator $A$ in the action is
the negative of the $D$-dimensional Laplacian $\Delta$ plus a constant
potential in the Euclidean space and the difference in the two models
is in the boundary conditions: Dirichlet boundary conditions on the
boundary of the cube and periodic boundary condition for the closed
domain. A complete system of eigenvalues is then:
$\lambda_{n,k}(T,q,l)=(2\pi T n)^2+\frac{\pi^2}{l^2}|k|^2+q$, where
$n\in \mathbb{Z}$ is an integer and  $k=(k_1,\dots,k_D)$ is a positive
integral vector in $(\mathbb{N}_0)^D$ in the first case, but is any
integral vector in $\mathbb{Z}^D$ in the second one ($2l$ is the
length of the circles in the torus in the second case). The zeta
functions are
\[
\zeta_{\mathcal{B}_D}(s;2\pi T,q,l)
=\sum_{(n,k)\in \mathbb{Z}\times(\mathbb{N}_0)^D}
\left[(2\pi T n)^2+\frac{\pi^2}{l^2}|k|^2+q\right]^{-s},
\]
\[
\zeta_{\mathcal{T}_D}(s;2\pi T,q,l)
={\sum}'_{(n,k)\in \mathbb{Z}^{D+1}}
\left[(2\pi T n)^2+\frac{\pi^2}{l^2}|k|^2+q\right]^{-s},
\]
for $\R(s)>\frac{D+1}{2}$. Using the results of Section \ref{s1}, we
just have to study the associated geometric zeta functions, namely
\[
\hskip -1.3truecm\zeta_{\mathcal{B}_D}(s;0,q,l) =\sum_{k\in
(\mathbb{N}_0)^D} \left[\frac{\pi^2}{l^2}|k|^2+q\right]^{-s}, \hskip
.5truecm \zeta_{\mathcal{T}_D}(s;0,q,l) ={\sum_{k\in
\mathbb{Z}^{D}}}' \left[\frac{\pi^2}{l^2}|k|^2+q\right]^{-s},
\]
for $\R(s)>\frac{D}{2}$. This is the aim of this section, that is
subdivided in three parts. In the first we introduce a
multidimensional generalization of the Riemann zeta function useful to
describe the geometric zeta functions, subsequently we calculate the
main zeta invariants, and in the  last part we give the partition
function and the thermodynamic functions. From now on we will assume
$D\geq 1$ if not otherwise stated.

\subsection{Multidimensional quadratic zeta
functions}\label{s31} Let $q^2$ be a complex constant that is not real
and negative, $n$ a $D$-dimensional vector with integer components in
$\mathbb{Z}^D$, and $A$ a real symmetric matrix of rank $D$ with
positive definite associated quadratic form, then we can introduce the
functions:
\[
\xi_D(s;q)=\sum_{n\in(\mathbb{N}_0)^D}(|n|^2+q^2)^{-s},
\]
\[
\zeta_D(s;A,q)
=\sum_{n\in\mathbb{Z}^D}(n^TA n+q^2)^{-s}
=q^{-2s}+\hat\zeta_D(s;n^T A n,q)
\]
when $\R(s)>\frac{D}{2}$. Notice that the definition introduced for
$\zeta_{D}(s;A,q)$ is ad hoc to avoid problems for the homogeneous
case $q=0$. In fact, while the definition given for $\xi_D(s;q)$
extends to $q=0$, in the Eisenstein sum one must omit the null vector
when $q=0$. With the definition above, this simply means to omit the
$q^{-2s}$ term in the homogeneous case. We will use the notation
$\zeta_D(s;q)$ for $\zeta_D(s;I,q)$ in the following. Notice in
particular that $\zeta_0(s;q)=q^{-2s}$ and
$2\xi_1(s;0)=\hat\zeta_1(s;0)=2\zeta_R(2s)$. Actually, the results we
are going to give for the Epstein zeta functions hold true for a large
class of zeta functions, that we introduce now. Let $n\in
\mathbb{Z}^D$ and $A$ be as above, $b$ and $x=(x_i)$ be real $D$
dimensional vectors, and assume $0\leq x_i<1$. Then, we define the
functions:
\[
\zeta_D(s;A,b,x,q)=\sum_{n\in \mathbb{Z}^D} [(n+x)^T A (n+x)+b^T n +q^2]^{-s},
\]
\[
\hat\zeta_D(s;A,b,x,q)=\sum_{n\in \mathbb{Z}^D_0}
[(n+x)^T A (n+x)+b^T n +q^2]^{-s},
\]
when $\R(s)>\frac{D}{2}$. Notice that $q^2$ must be non vanishing in
the definition of $\zeta_D(s;A,b,x,q)$. A lot is known about these
multidimensional zeta functions, in particular the homogeneous case,
namely the Epstein zeta function, has been deeply investigated (see
\cite{Ter} for a good overview, or locally cited references). Here, we
collect a series of results that seem more interesting and useful for
the present purposes. As we will see, it is easier to get more general
results for the Epstein series, and this is essentially due to the
possibility of more effective use of the Poisson summation formula.
All the proofs are based on classical techniques, namely the Mellin
transform and the Poisson summation formula; since these tools were
used for all the proofs in the previous sections we omit to give
details here and refer the interested reader to the literature
available for a deeper account on this subject.

\noindent We begin by introducing some analytic representations. These
will be useful to get all information about the analytic extensions of
the zeta functions, as well as when calculations are involved to
evaluate them at some particular value.
\begin{lem}\hskip .5truecm
$ \xi_D(s;q) =-\frac{1}{2}\xi_{D-1}(s;q)
+\frac{\sqrt{\pi}}{\Gamma(s)}\Gamma\left(s-\frac{1}{2}\right)
\xi_{D-1}\left(s-\frac{1}{2};q\right)+$
\[
+ \frac{4\pi^s}{\Gamma(s)}\sum_{(n,k)\in(\mathbb{N}_0)^{D}}
\left(\frac{\sqrt{|k|^2+q^2}}{n}\right)^{\frac{1}{2}-s}
K_{s-\frac{1}{2}}(2\pi n \sqrt{|k|^2+q^2}).
\]
\label{l311}
\end{lem}
For the Epstein type functions, we have the following lemma when
$q\not= 0$:
\begin{lem}$~$
\scriptsize
\[
\hskip -2truecm
\zeta_D(s;A,0,x,q)=\frac{\pi^\frac{D}{2}}{\sqrt{{\rm det}A}}
\frac{\Gamma\left(s-\frac{D}{2}\right)}{\Gamma(s)}q^{-2s+D}
+\frac{2\pi^s}{\sqrt{{\rm det}A}\Gamma(s)}q^{\frac{D}{2}-s}
{\sum_{n\in \mathbb{Z}_0^D}} \left(\frac{\sqrt{n^T A^{-1}
n}}{q}\right)^{s-\frac{D}{2}} K_{s-\frac{D}{2}}(2\pi q\sqrt{n^T
A^{-1}n}),
\]
\[
\hskip -2truecm \zeta_D(s;A,b,0,q)=\frac{\pi^\frac{D}{2}}{\sqrt{{\rm
det}A}} \frac{\Gamma\left(s-\frac{D}{2}\right)}{\Gamma(s)}
\left(q^2-\frac{1}{4}b^TA^{-1}B\right)^{\frac{D}{2}-s}+
\]
\[
\hskip -2truecm +\frac{2\pi^s}{\sqrt{{\rm det}A}\Gamma(s)} {\sum_{n\in
\mathbb{Z}_0^D}} \left(\frac{n^T A^{-1}
n}{q^2-\frac{1}{4}b^TA^{-1}b}\right)^
{\frac{1}{2}\left(s-\frac{D}{2}\right)} K_{s-\frac{D}{2}}(\pi
\sqrt{4q^2-b^TA^{-1}b}\sqrt{n^T A^{-1}n}),~{\rm if}~4q^2-b^TA^{-1}b>0.
\]
\end{lem}
When $q=0$, we need some more notation. Let $a_{i,j}$ be the elements
of $A$. Let $A_{1,1}$ the minor of $a_{1,1}$ in $A$, and $A_1$ denotes
the $D-1$-column vector whose elements are the elements of the first
line of $A_{1,1}$. Let $\hat b_1$ and $\hat x_1$ be the $D-1$ vectors
whose elements are the last $D-1$ elements of $b$ and $x$,
respectively. Let $B$ be the $D-1$ square matrix whose elements are
$b_{i-1,j-1}=a_{i,j}-\frac{a_{1,i}a_{1,j}}{a_{1,1}}$, where the
indices $i$ and $j$ run from $2$ to $D-1$.
\begin{lem}\scriptsize
\hskip .5truecm $\hat\zeta_D(s;A,0,x,0)=\hat\zeta_1(s;a_{1,1},0,x_1,0)+
\frac{\sqrt{\pi}\Gamma\left(s-\frac{1}{2}\right)}
{\sqrt{a_{1,1}}\Gamma(s)} \hat\zeta_{D-1}\left(s-\frac{1}{2};B,0,\hat
x_1,0\right)+\frac{4\pi^s}{\Gamma(s)}
a_{1,1}^{-\frac{1}{2}\left(s-\frac{1}{2}\right)} \times$
\[
\times \sum_{n_1=1}^\infty\sum_{n\in \mathbb{Z}_0^{D-1}} \cos
2\pi\left[n_1 x_1+\frac{1}{a_{1,1}}A_1(n+\hat x_1) n_1\right]
\frac{n_1^{s-\frac{1}{2}}}
{(n^TBn)^{\frac{1}{2}\left(s-\frac{1}{2}\right)}}
K_{s-\frac{1}{2}}\left(\frac{2\pi n_1}{\sqrt{a_{1,1}}} \sqrt{n^T B
n}\right),
\]
\hskip .9truecm$ \hat\zeta_D(s;A,b,0,0) =\hat\zeta_1(s;a_{1,1},b_1,0,0)
+\frac{\sqrt{\pi}\Gamma\left(s-\frac{1}{2}\right)}{\sqrt{a_{1,1}}\Gamma(s)}
\hat\zeta_{D-1}
\left(s-\frac{1}{2};B,0,\frac{b_1}{a_{1,1}}A_1,-\frac{b_1^2}{4a_{1,1}}\right)+
\frac{4\pi^s}{\Gamma(s)}
a_{1,1}^{-\frac{1}{2}\left(s-\frac{1}{2}\right)}\times $

$\times \sum_{n_1=1}^\infty\sum_{n\in \mathbb{Z}_0^{D-1}} \cos
\left[\frac{\pi}{2a_{1,1}}\left(2A_1n+b_1\right)n_1\right]
\frac{n_1^{s-\frac{1}{2}} K_{s-\frac{1}{2}}\left(\frac{2\pi n_1
}{\sqrt{a_{1,1}}} \sqrt{n^T A_{1,1} n -\frac{1}{4a_{1,1}}(2A_1
n+b_1)^2}\right)} {\left(n^TA_{1,1}n-\frac{1}{4a_{1,1}}(2A_1
n+b_1)^2\right) ^{\frac{1}{2}\left(s-\frac{1}{2}\right)}}= $

\hskip -.7truecm $\hat\zeta_{D-1}(s;A_{1,1},\hat b_1,0,0)
+\frac{\pi^\frac{D-1}{2}\Gamma\left(s-\frac{D-1}{2}\right)}
{\sqrt{{\rm det}A_{1,1}}\Gamma(s)}
\hat\zeta_1\left(s-\frac{D-1}{2};\frac{{\rm det}A}{{\rm det}A_{1,1}},
b_1-A_1^T A_{1,1}^{-1} \hat b_1,0, -\frac{1}{4}\hat b_1^T
A_{1,1}^{-1}\hat b_1\right) +\frac{2\pi^2}{\sqrt{{\rm
det}A_{1,1}}}\times$

\hskip -.7truecm $ \sum_{n\in\mathbb{Z}_0^{D-1}} \e^{\pi i n^T
A_{1,1}^{-1}(2n_1 A_1+\hat b_1)} (n^T A_{1,1}^{-1}
n)^{\frac{1}{2}\left(s-\frac{D-1}{2}\right)}
\frac{K_{s-\frac{D-1}{2}}\left(2\pi \sqrt{n^T A_{1,1}^{-1} n}
\sqrt{\frac{{\rm det}A}{{\rm det}A_{1,1}}n_1^2 +(b_1-A_1^T
A_{1,1}^{-1} \hat b_1)n_1 -\frac{1}{4}\hat b_1^T A_{1,1}^{-1}\hat
b_1}\right)} {\left(\frac{{\rm det}A}{{\rm det}A_{1,1}}n_1^2
+(b_1-A_1^T A_{1,1}^{-1} \hat b_1)n_1 -\frac{1}{4}\hat b_1^T
A_{1,1}^{-1}\hat b_1\right)
^{\frac{1}{2}\left(s-\frac{D-1}{2}\right)}}, $ \normalsize \\where the
last two representations hold if $-\frac{b^2_1}{4a_{1,1}}$ or
$-\frac{1}{4} \hat b_1^T A_{1,1}^{-1} \hat b_1$ are not negative
integers. \label{l312}
\end{lem}
More Chowla-Selberg type formulas \cite{CS} \cite{BG} can be
found in \cite{Ter}. We also recall the important reflection
formula \cite{Sei} \cite{Tay}
\begin{lem}\hskip 1truecm
$ \pi^{-s}\Gamma(s)\hat\zeta_D(s;A,0,0,0)
=\frac{\pi^{s-\frac{D}{2}}}{\sqrt{{\rm det}A}}
\Gamma\left(\frac{D}{2}-s\right)
\hat\zeta_D\left(\frac{D}{2}-s;A^{-1},0,0,0\right). $
 \label{lll}
\end{lem}
\begin{corol}$
\hskip 1truecm \zeta_D'(0;A,0,0,q) =\left\{\begin{array}{cc}
\frac{\pi^\frac{D}{2}}{\sqrt{\det A}}
\Gamma\left(-\frac{D}{2}\right)q^D&D~{\rm odd}\\
\frac{\pi^\frac{D}{2}}{\sqrt{\det
A}}\frac{(-1)^\frac{D}{2}}{\frac{D}{2}!} q^D&D~{\rm
even}\end{array}\right.+$
\[
\hskip 1truecm+ \frac{2\pi^{\frac{D}{2}}}{\sqrt{\det A}} {\sum_{n\in
\mathbb{Z}^D_0}} (n^T A^{-1} n)^{-\frac{D}{4}} K_{-\frac{D}{2}}(2\pi
q\sqrt{n^T A^{-1}n}).
\]
\label{cp32a}
\end{corol}
See also \cite{CS} and \cite{Ter} for the Kronecker limit formula.
\begin{corol}  For $n=1,2,3,\dots$: $ \hat\zeta_D(-n;A,0,0,0)=0$;
for $n=0,1,2,\dots$, $q\not= 0$
\[
\hskip -2truecm \zeta_D(-n;A,0,0,q)=
q^{2n}+\hat\zeta_D(-n;A,0,0,q)=\left\{\begin{array}{cc}0&D~{\rm odd}\\
\frac{(-1)^{-n}\frac{D}{2}!}{\left(n+\frac{D}{2}\right)!}
\frac{\pi^\frac{D}{2}}{\sqrt{{\rm det} A}} q^{2n+D}&D~{\rm even};
\end{array}\right.
\]
\label{cp32b}
\end{corol}
Eventually, using the representation introduced in the previous
lemmas, or using classical methods, we get all information about
poles, residua and particular values.
\begin{lem} The function $\xi_D(s;q)$ extends analytically to the whole
complex plane up to simple poles at
$s=\frac{D}{2},\frac{D-1}{2},\dots, \frac{1}{2}$ and
$s=-\frac{1}{2}-j$, $j=0,1,2,\dots$, with residua: $ {\rm
Res}_1\left(\xi_D(s;q),-\frac{k}{2}\right)
=\frac{(-1)^D}{2^D\Gamma\left(-\frac{k}{2}\right)}
\sum_{j=1}^D\sum_{i= 0, 2i=j+k}^\infty \frac{(-1)^{i+j}}{i!}
\left(\begin{array}{c} D\\j\end{array}\right) \pi^\frac{j}{2}q^{2i}$,
$k=-D,-D+1,\dots,$ $k\notin 2\mathbb{N}$;
$\xi_D(0;q)=\frac{(-1)^D}{2^D}
+\frac{(-1)^D}{2^D}\sum_{j=1}^\frac{D}{2} \left(\begin{array}{c}
D\\2j\end{array}\right) \frac{(-1)^j \pi^j q^{2j}}{\Gamma(j+1)}$.
\label{l313}
\end{lem}
Notice that in the homogeneous case there are poles at
$s=\frac{1}{2},\dots, \frac{D}{2}$, with residuum $ {\rm
Res}_1\left(\xi_D(s;0),s=\frac{i}{2}\right)
=\frac{(-1)^{D+i}}{2^D\Gamma\left(\frac{i}{2}\right)}
\left(\begin{array}{c}D\\i\end{array}\right) \pi^\frac{i}{2}$,
$i=1,2,\dots ,D$, $ \xi_D(0;0)=\frac{(-1)^D}{2^D}$.

\begin{lem} The analytic continuation
of $\hat\zeta_D(s;A,0,0,q)$ and $\zeta_D(s;A,0,0,q)$ are regular on
the whole complex $s$-plane up to a simple poles at $s=\frac{D}{2}-j$,
$j=0,1,2,\dots$ if $D$ is odd and $s=\frac{D}{2}-j$,
$j=0,1,2,\dots,\frac{D}{2}-1$ if $D$ is even, respectively. The
residua are, for both the functions, $ {\rm
Res}_1\left(\hat\zeta_D(s;A,0,0,q),s=\frac{D}{2}-j\right)
=\frac{(-1)^j}{j!} \frac{\pi^\frac{D}{2}q^{2j}} {\sqrt{{\rm
det}A}\Gamma\left(\frac{D}{2}-j\right)}$, $j$ as before. Moreover:
$\zeta_D(0;A,0,0,q)=1+\hat\zeta(0;A,0,0,q)
=\left\{\begin{array}{cc}0&D~{\rm odd}\\
\frac{(-\pi)^\frac{D}{2} q^D}{\frac{D}{2}!}&D~{\rm even}
\end{array}\right.$.
\label{l314}
\end{lem}
Notice that in the homogeneous case the unique pole is at
$s=\frac{D}{2}$ with residuum $\frac{\pi^\frac{D}{2}}{\sqrt{{\rm
det}A}\Gamma\left(\frac{D}{2}\right)}$, and
$\hat\zeta(0;A,0,0,0)=-1$ for all $D$.

\subsection{Dirichlet boundary conditions}
\label{s32} In this case, the spectrum of $-\Delta_M$ is positive
definite, and we can write the geometric zeta function using the
function $\xi_D(s;q)$ just introduced:
$\zeta_{\mathcal{B}_D}(s;0,q,l)
=\frac{l^{2s}}{\pi^{2s}}\xi_D\left(s;\frac{\sqrt{q}l}{\pi}\right)$.
Notice that this  extends continuously to the homogeneous case
$q=0$. Using Propositions \ref{p1b} and \ref{p2} in Section
\ref{s1}, and Lemma \ref{l313} of the previous part, we get:
\begin{prop} The function $\zeta_{\mathcal{B}_D}(s;y,q,l)$
extends analytically to a regular function on the whole complex
$s$-plane up to simple poles for all the $s=\frac{D+1-j}{2}$, with
$j=0,1,2,\dots$, that are not non positive integers, with residua
($k=-D-1, -D, -D+1,\dots, k\notin \mathbb{N}$):
\[
\hskip -2.3truecm{\rm
Res}_1\left(\zeta_{\mathcal{B}_D}(s;y,q,l),s=-\frac{k}{2}\right)
=\frac{(-1)^D \pi^{k+1}}{2^D l^{k+1}\Gamma\left(-\frac{k}{2}\right) y}
\sum_{i=0}^D\sum_{j=0,2j-i-1=k}^\infty
\left(\begin{array}{c}D\\i\end{array}\right)
\frac{(-1)^{i+j}l^{2j}q^{j}}{j!\pi^{2j-\frac{i+1}{2}}}.
\]
\label{p31}
\end{prop}
Notice that in the homogeneous case the poles  are at
$s=\frac{D+1}{2},\frac{D}{2},\dots \frac{1}{2}$.
\begin{prop} For all $y>0$, fixed $q\geq 0$ and $l> 0$, and uniformly in $s$ near $s=0$,
\[
\zeta_{\mathcal{B}_D}(s;y,q,l)
=\frac{l^{2s-1}}{\pi^{2s-1}}
\frac{\sqrt{\pi}}{\Gamma(s)}y^{-1}
\Gamma\left(s-\frac{1}{2}\right)
\xi_D\left(s-\frac{1}{2};\frac{\sqrt{q}l}{\pi}\right)+
\]
\[
+\frac{4\pi^s}{\Gamma(s)} y^{-s-\frac{1}{2}} \sum_{(n,k)\in
(\mathbb{N}_0)^{D+1}}
\left(\frac{n}{\sqrt{\frac{\pi^2}{l^2}|k|^2+q}}\right)^{s-\frac{1}{2}}
K_{s-\frac{1}{2}} \left(\frac{2\pi
n}{y}\sqrt{\frac{\pi^2}{l^2}|k|^2+q}\right).
\]
\label{p6}
\end{prop}
As a corollary, we give the formula for the derivative at $s=0$.
Some care is necessary to deal with the first term when $q\not=
0$. In this case in fact, the function $\xi_D(s;q)$ has a pole at
$s=-\frac{1}{2}$. Proceeding as in the proof of the corollary to
Proposition \ref{p2}, we get
\begin{corol} For all fixed $q\geq 0$ and $l> 0$,
\[
\zeta_{\mathcal{B}_D}(0;y,q,l)=-\frac{2\pi^2}{l}  R_1 \frac{1}{y}
=\frac{(-1)^D}{2^D y} \sum_{i=0}^{\left[\frac{D-1}{2}\right]}
\left(\begin{array}{c}D\\2i+1\end{array}\right)
\frac{(-1)^{i}}{(i+1)!} \frac{l^{2i+1}q^{i+1}}{\pi^i},
\]
\[
\zeta_{\mathcal{B}_D}'(0;y,q,l) =A_D(q,l)\frac{2\pi}{y}
-2\log \prod_{k\in (\mathbb{N}_0)^D}
\left(1-\e^{-\frac{2\pi}{y}\sqrt{\frac{\pi^2}{l^2}|k|^2+q}}\right),
\]
where $A_D(q,l) =-\frac{\pi}{l}\left[R_0+(2-\log 4\pi+\log
l)R_1\right]$, $ R_i={\rm Res}_i
\left(\xi_D\left(s;\frac{\sqrt{q}l}{\pi}\right),s=-\frac{1}{2}\right)$.
\label{c6}
\end{corol} Notice that in the homogeneous case the
value at $s=0$ is $0$ for all $D$. Also, when $q=0$, $\xi_D(s;0)$
is regular at $s=-\frac{1}{2}$; thus, an explicit formula for the
constant when $q=0$ is $ A_{D}(0,l)
=-\frac{\pi}{l}\xi_D\left(-\frac{1}{2};0\right)$, but it is more
complicate otherwise. Using Lemma \ref{l311}, if $q=0$, when $D=1$
we get twice the Riemann zeta function at $s=-1$, while for
higher $D$ only numerical evaluations are possible and give: $
\xi_1\left(-\frac{1}{2};0\right)=\zeta_R(-1)=-\frac{1}{12},\hskip
.05in \xi_2\left(-\frac{1}{2};0\right)=0.026127,\hskip .05in
\xi_3\left(-\frac{1}{2};0\right)=-0.010015$. Notice that these
results can be easily obtained using the representations
introduced in Lemma \ref{l311}, since the series there converges
very fast. The same computation can be made using the reflection
formula for the multidimensional zeta function given in Lemma
\ref{lll}, but in that case the series converges very slowly and
a much longer computation is necessary.

\noindent We conclude this part with some remarks on the well know
cases $D=0$ and $D=1$. When $D=0$ \cite{Spr4},
\[
z_1(s;y,q)=\zeta_{\mathcal{B}_0}(s;y,q,\pi)
=\sum_{n\in\mathbb{Z}}(y^2n^2+q)^{-s}
=q^{-s}+2y^{-2s}\xi_1\left(s;\frac{\sqrt{q}}{y}\right);
\]
for $q\not=0$, $z_1(s;y,q)$ has a simple pole at
$s=\frac{1}{2}-j$, $j=0,1,\dots$, with residuum $\frac{(-1)^j}{j!}
\frac{\sqrt{\pi}q^{j}}{\Gamma\left(\frac{1}{2}-j\right) y}$, and
$z_1(0;y,q)=0$, $z_1'(0,y;q)=-2\log 2{\rm sh}\frac{\pi
\sqrt{q}}{y}$. For $q=0$, $z_1(s;y,0))=2y^{-2s}\zeta_R(2s)$,  has
a single pole at $s=\frac{1}{2}$ with residuum $\frac{2}{y}$, and $
z_1(0;y,0)=-1$, $z'(0;y,0)=2\log y-2\log 2\pi$.

\noindent When $D=1$, we have the zeta function associate to the
Laplacian plus a constant potential on a cylinder \cite{Wei}
\cite{Spr4},
\[
z_2(s;y,q)=\zeta_{\mathcal{B}_1}(s;y,q,\pi)
=\sum_{n\in\mathbb{Z}}\sum_{k=1}^\infty
(y^2n^2+k^2+q)^{-s},
\]
and $ z_2(0;y,q)=-\frac{\pi q}{2y}$, $z'_2(0;y,q)=
\frac{\pi}{6}\frac{1}{y}-(\gamma-\log 2)\frac{\pi q}{y}
-2\log\prod_{n=1}^\infty
\left(1-\e^{-\frac{2\pi}{y}\sqrt{n^2+q^2}}\right)
-\frac{2\pi}{y}\sum_{j=2}^\infty
\left(\begin{array}{c}\frac{1}{2}\\j\end{array} \right)
\zeta_R(2j-1)q^j$. In particular, when $q=0$, the first term of
the $s$-expansion near $s=0$ of such zeta function is well known
\[
z'_2(0;y,0)
=\frac{\pi}{6}\frac{1}{y}
-2\log\prod_{n=1}^\infty \left(1-\e^{-\frac{2\pi}{y} n}\right)
=-2\log\eta\left(\frac{i}{y}\right),
\]
where $\eta(z)$ is the Dedekind eta function. In such case, it is
very easy to pass from the small $y$ to the large $y$ expansion
using the well known modular transformation of the eta function,
namely
$\eta\left(-\frac{1}{\tau}\right)=\sqrt{\frac{\tau}{i}}\eta(\tau)$.
It is also easy to see how the presence of a non homogeneous term
breaks this symmetry. A deeper investigation about this point is
performed in the next part.

\noindent A final observation is about the large $y$ expansion. As just
noticed, this can be immediately worked out when $D=1$. For higher
values of $D$, it is harder. In fact, we can not use the analytic
representation given in Proposition \ref{p6} for large $y$ neither we
know how the zeta function behaves under the modular transformation
$y\to\frac{1}{y}$. Despite that, a direct approach is still possible
for fixed $D$, consisting in expressing the zeta function in dimension
$D$ recursively in terms of the zeta function in dimension $D-1$, and
using the known behavior in dimension $D=1$. Calculations are tedious
but straightforward; the leading term in the expansion is consistent
with the one obtained using Proposition \ref{p4}, but now we can get
further terms.

\subsection{Periodic boundary conditions}
\label{s33} Recall that in the present case the spectrum of
$-\Delta_M$ is not always positive definite; more precisely,
$\mathcal{K}_q=1$ if $q=0$, but it vanishes if $q>0$. We need
the functions $\zeta_D$ and $\hat\zeta_D$ of  \ref{s31} in order to
write the geometric zeta function: $ \zeta_{\mathcal{T}_D}(s;0,q,l)
=\frac{l^{2s}}{\pi^{2s}}\zeta_D\left(s;\frac{\sqrt{q}l}{\pi}\right)
=q^{-2s}
+\frac{l^{2s}}{\pi^{2s}}\hat\zeta_D\left(s;\frac{\sqrt{q}l}{\pi}\right)$.
Due to the presence of the first term, it is clear how this expression
does not extend to the homogeneous case. In the present situation it
is easier to deal with the two cases independently. This does not
affect the poles, hence we can state the following unique result using
Propositions \ref{p1b} and \ref{p2} in Section \ref{s1}, and Lemma
\ref{l312} of \ref{s31}.
\begin{prop} The functions $\zeta_{\mathcal{T}_D}(s;y,q,l)$
and $\hat\zeta_{\mathcal{T}_D}(s;y,q,l)$ extend analytically to a
regular function on the whole complex $s$-plane up to simple poles for
all the $s=\frac{D+1}{2}-j$, with $j=0,1,2,\dots$, if $D$ is even, and
$j=0,1,2,\dots, \frac{D+1}{2}-1$, if $D$ is odd. The residua are for
both the functions: $ {\rm
Res}_1\left(\zeta_{\mathcal{T}_D}(s;y,q,l),s=\frac{D+1}{2}-j\right)
=\frac{(-q)^j}{j!\Gamma\left(\frac{D+1}{2}-j\right)}
\frac{l^D}{\pi^\frac{D-1}{2}}\frac{1}{y}$.
\label{p32}
\end{prop}
Notice that in the homogeneous case, the unique pole is at
$s=\frac{D+1}{2}$ with residuum
$\frac{l^D}{\pi^\frac{D-1}{2}\Gamma\left(\frac{D+1}{2}-j\right)}\frac{1}{y}$.
An analytic representation analogous to the one stated in
Proposition \ref{p6} of the previous part is
\begin{prop} For all $y>0$, $l>0$, and uniformly in $s$ near $s=0$,
\[
\zeta_{\mathcal{T}_D}(s;y,q,l)
=\frac{l^{2s-1}}{\pi^{2s-1}}
\frac{\sqrt{\pi}}{\Gamma(s)}y^{-1}
\Gamma\left(s-\frac{1}{2}\right)
\zeta_D\left(s-\frac{1}{2};\frac{\sqrt{q}l}{\pi}\right)+
\]
\[
+\frac{4\pi^s}{\Gamma(s)} y^{-s-\frac{1}{2}}
\sum_{n=1}^\infty \sum_{k\in \mathbb{Z}^D}
\left(\frac{n}{\sqrt{\frac{\pi^2}{l^2}|k|^2+q}}\right)^{s-\frac{1}{2}}
K_{s-\frac{1}{2}}\left(\frac{2\pi n}{y}\sqrt{\frac{\pi^2}{l^2}|k|^2+q}\right),
\]
uniformly in $q$ for $q$ in any closed subset of the positive
real axis, while
\[
\zeta_{\mathcal{T}_D}(s;y,0,l)
=\frac{l^{2s-1}}{\pi^{2s-1}}
\frac{\sqrt{\pi}}{\Gamma(s)}y^{-1}
\Gamma\left(s-\frac{1}{2}\right)\hat\zeta_D\left(s-\frac{1}{2};0\right)
+2 y^{-2s}\zeta_R(2s)+
\]
\[
+\frac{4\sqrt{\pi}l^{s-\frac{1}{2}}}{\Gamma(s)} y^{-s-\frac{1}{2}}
\sum_{n=1}^\infty \sum_{k\in \mathbb{Z}_0^D}
\left(\frac{n}{|k|}\right)^{s-\frac{1}{2}}
K_{s-\frac{1}{2}}\left(\frac{2\pi^2 n}{ly}|k|\right).
\]
\label{p1}
\end{prop}
\begin{corol} Uniformly in $q>q_0>0$:
\[
\zeta_{\mathcal{T}_D}(0;y,q,l)=-\frac{2\pi^2}{l} R_1 \frac{1}{y}
=\left\{\begin{array}{ll} 0 & D~{\rm even}\\
\frac{(-q)^\frac{D+1}{2} l^D}{\frac{D+1}{2}! \pi^\frac{D-1}{2}}\frac{1}{y}
&D~{\rm odd},\end{array}\right.
\]
\[
\zeta_{\mathcal{T}_D}'(0;y,q,l) =B_D(q,l)\frac{2\pi}{y}
-2\log \prod_{k\in \mathbb{Z}^D}
\left(1-\e^{-\frac{2\pi}{y}\sqrt{\frac{\pi^2}{l^2}|k|^2+q}}\right);
\]
\[
\zeta_{\mathcal{T}_D}(0;y,0,l)=-\frac{2\pi^2}{l} \hat R_1
\frac{1}{y}+2 \zeta_R(0)=-1,
\]
\[
\zeta_{\mathcal{T}_D}'(0;y,0,l)
=\hat B_D(0,l)\frac{2\pi}{y}
+2\log\frac{y}{2\pi}
-2\log \prod_{k\in \mathbb{Z}_0^D}
\left(1-\e^{-\frac{2\pi^2}{ly}|k|}\right),
\]
where $B_D(q,l)=-\frac{\pi}{l}\left[R_0+(2-\log 4\pi +\log
l)R_1\right]$, $ R_i={\rm Res}_i
\left(\zeta_D\left(s;\frac{\sqrt{q}l}{\pi}\right),s=-\frac{1}{2}\right)$,
and similarly for the hatted ones. \label{c7}
\end{corol}
We get simple expressions for the constants when $D$ is even or
when $q=0$. In particular, in the second case, $\hat B_{D}(0,l)
=-\frac{\pi}{l}\hat\zeta_D\left(-\frac{1}{2};0\right)$. Using
Lemma \ref{l312} and numerical evaluations: $
\hat\zeta_1\left(-\frac{1}{2};0\right)=2\zeta_R(-1)=-\frac{1}{6},\hskip
.05in \hat\zeta_2\left(-\frac{1}{2};0\right)=-0.2286,\hskip .05in
\hat\zeta_3\left(-\frac{1}{2};0\right)=-0.26493$.

\noindent Of particular interest is the case $D=1$. As stated in the
previous part, this is related with the Dedekind eta function,
$\eta(z)$. Namely, assuming $l=\pi$ for simplicity,
\[
\lim_{q\to
0^+}\left[\zeta_{\mathcal{T}_1}'(0;y,q,\pi)+\log q\right]
=2\log y-2\log2\pi -4\log\eta\left(\frac{i}{y}\right).
\]
This suggest to define the function
\[
\eta(\tau,q)=-\e^{\pi i \tau B_2(q)}\left(1-\e^{2\pi i \tau q}\right)
\prod_{n=1}^\infty\left(1-\e^{2\pi i\tau \sqrt{n^2+q^2}}\right)^2,
\]
for real positive $q$ and complex $\tau$ with positive imaginary part.
It is easy to check that $ \lim_{q\to 0^+} \frac{\eta(\tau,q)}{2\pi
i\tau q}=\eta^2(\tau)$. It is also easy to realize that the presence
of the non homogeneous term $q$, breaks modularity. On the other side,
the modular transformation for the Dedekind eta function can be
deduced using the symmetry in the definition of the zeta function
$\zeta_{\mathcal{T}_1}(s;y,q^2,\pi)$ under the exchange of the
summation indices in the first term of the $s$-expansion near $s=0$.
Using the same symmetry for the function $\eta(iy,q)$, we get instead
of the modular transformation the following relation
\scriptsize
\[
\hskip -1truecm \log\eta\left(\frac{i}{y},qy\right) =\log\eta(iy,q)-\pi
q^2y \log y +2\pi y \sum_{j=2}^\infty
\left(\begin{array}{c}\frac{1}{2}\\j\end{array}\right) \zeta_R(2j-1)
q^{2j} +\frac{2\pi}{y} \sum_{j=2}^\infty
\left(\begin{array}{c}\frac{1}{2}\\j\end{array}\right) \zeta_R(2j-1)
(qy)^{2j}.
\]
\normalsize Since the behavior of $\eta(iy,q)$ for large $y$ is clear,
the above expression can be used (exactly as it was for the Dedekind
zeta function) to deduce the behavior for small $y$. We get, for $y\to
0^+$: \scriptsize
\[
\hskip -1truecm\eta(iy,q)=-\frac{\pi}{6y}+\pi q^2 y\log y+\pi(q+1)
-\left[ \sum_{j=2}^\infty
\left(\begin{array}{c}\frac{1}{2}\\j\end{array}\right) \zeta_R(2j-1)
q^{2j}+\pi q^2\right] y-\frac{\pi y^3 }{4}\zeta_R(3)+O(y^4).
\]
\normalsize

\subsection{Thermodynamic functions} \label{s34} We write now explicit
formulas for the partition function of the models introduced in
\ref{s31}. Such formulas can be used to get explicit expressions
for all the thermodynamic functions. In particular, the behavior
for low and high temperature are given. The partition function for
a massive scalar thermal radiation at temperature $T$ in a box of
volume $l^D$ and on the torus $\mathcal{T}_D$ are, for any fixed
positive $l$,
\[
\log Z_{\mathcal{B}_D}(T,q,l)
=\frac{1}{2}A_D(q,l)\frac{1}{T}-\log\prod_{k\in(\mathbb{N}_0)^D}
\left(1-\e^{-\frac{1}{T}\sqrt{\frac{\pi^2}{l^2}|k|^2+q}}\right)+
\]
$$
+\frac{(-1)^{D+1}}{2^{D+2}} \log \rho \frac{1}{T}
\sum_{i=0}^{\left[\frac{D-1}{2}\right]}
\left(\begin{array}{c}D\\2i+1\end{array}\right)
\frac{(-1)^{i}}{(i+1)!} \frac{l^{2i+1}q^{i+1}}{\pi^i},
$$
\[
\log Z_{\mathcal{T}_D}(T,q,l) =\frac{1}{2}B_D(q,l)\frac{1}{T}
-\log\prod_{k\in\mathbb{Z}^D}
\left(1-\e^{-\frac{1}{T}\sqrt{\frac{\pi^2}{l^2}|k|^2+q}}\right)+
\]
$$
-\frac{1}{2}\log\rho \frac{1}{T}\left\{\begin{array}{ll} 0 & D~{\rm even}\\
\frac{(-q)^\frac{D+1}{2} l^D}{2\frac{D+1}{2}! \pi^\frac{D+1}{2}}
&D~{\rm odd},\end{array}\right.
$$
these are smooth functions of the temperature $T$, for bounded
$T\geq 0$, uniformly in the mass term $q$, for $q$ in any
closed interval of the positive real axis. When $q=0$,
\[
\hskip -2cm \log Z_{\mathcal{B}_D}(T,0,l)
=-\frac{\pi}{2l}\xi_D\left(-\frac{1}{2};0\right)\frac{1}{T}
-\log\prod_{k\in(\mathbb{N}_0)^D}\left(1-\e^{-\frac{\pi}{lT}|k|}\right),
\]
\[
\hskip -2cm \log Z_{\mathcal{T}_D}(T,0,l)
=-\frac{\pi}{2l}\hat\zeta\left(-\frac{1}{2};0\right)\frac{1}{T}
+\log T
-\log\prod_{k\in\mathbb{Z}_0^D}\left(1-\e^{-\frac{\pi}{lT}|k|}\right)+\frac{1}{2}\log
\rho,
\]
where some values for the multidimensional Riemann zeta functions are
given in \ref{s32} and \ref{s33} respectively. Using these
expressions, we get the behavior of the main thermodynamic functions:
partition function, energy, entropy, pressure of the radiation, and
specific heat. For low $T$, fixed $q$ and $l$, we get on the box
$$
\hspace{-6cm}\log Z_{\mathcal{B}_D}(T,q,l)
=\frac{1}{2}A_D(q,l)\frac{1}{T}+
$$
$$
\hspace{2cm} +\frac{(-1)^{D+1}}{2^{D+2}} \log \rho \frac{1}{T}
\sum_{i=0}^{\left[\frac{D-1}{2}\right]}
\left(\begin{array}{c}D\\2i+1\end{array}\right)
\frac{(-1)^{i}}{(i+1)!} \frac{l^{2i+1}q^{i+1}}{\pi^i}
+O(\e^{-\frac{1}{T}}),
$$
$$
F_{\mathcal{B}_D}(T,q,l)=-\frac{1}{2}A_D(q,l)
+\frac{(-1)^{D}}{2^{D+2}} \log \rho
\sum_{i=0}^{\left[\frac{D-1}{2}\right]}
\left(\begin{array}{c}D\\2i+1\end{array}\right)
\frac{(-1)^{i}}{(i+1)!} \frac{l^{2i+1}q^{i+1}}{\pi^i}
+O(T\e^{-\frac{1}{T}}),
$$
\[
S_{\mathcal{B}_D}(T,q,l)=O(T^{-1}\e^{-\frac{1}{T}}), \hskip 1truecm
c_{\mathcal{B}_D}(T,q,l)=O(T^{-2}\e^{-\frac{1}{T}}).
\]
while on the torus, we must distinguish the $q=0$ case:
$$
\log Z_{\mathcal{T}_D}(T,q,l)
=\frac{1}{2}B_D(q,l)\frac{1}{T}-\frac{1}{2}\log\rho\frac{1}{T}\left\{\begin{array}{ll} 0 & D~{\rm even}\\
\frac{(-q)^\frac{D+1}{2} l^D}{2\frac{D+1}{2}! \pi^\frac{D+1}{2}}
&D~{\rm odd}\end{array}\right. +O(\e^{-\frac{1}{T}}),
$$
$$
F_{\mathcal{T}_D}(T,q,l)=-\frac{1}{2}B_D(q,l)+\frac{1}{2}\log\rho\left\{\begin{array}{ll} 0 & D~{\rm even}\\
\frac{(-q)^\frac{D+1}{2} l^D}{2\frac{D+1}{2}! \pi^\frac{D+1}{2}}
&D~{\rm odd}\end{array}\right. +O(T\e^{-\frac{1}{T}}),
$$
$$
S_{\mathcal{T}_D}(T,q,l)=O(T^{-1}\e^{-\frac{1}{T}}), \hskip 1truecm
c_{\mathcal{T}_D}(T,q,l)=O(T^{-2}\e^{-\frac{1}{T}}),
$$
\[
\log Z_{\mathcal{T}_D}(T,0,l)
=-\frac{\pi}{2l}\hat\zeta\left(-\frac{1}{2};0\right)\frac{1}{T}+\log
T +\frac{1}{2}\log\rho+O(\e^{-\frac{1}{T}}),
\]
\[
F_{\mathcal{T}_D}(T,0,l) =-T\log
T+\hat\zeta\left(-\frac{1}{2};0\right)\frac{\pi}{2l}
-\frac{1}{2}T\log\rho +O(T\e^{-\frac{1}{T}}),
\]
\[
S_{\mathcal{T}_D}(T,0,l)=\log T+\frac{1}{2}\log
\rho+1+O(T^{-1}\e^{-\frac{1}{T}}),
\]
\[
c_{\mathcal{T}_D}(T,0,l)=1+O(T^{-2}\e^{-\frac{1}{T}}).
\]
Notice that the null mass case needs no independent treatment on
the box. Recalling the remark at the end of \ref{s32} or
the Proposition \ref{p4}, we get the behaviors for
high $T$, fixed $q$ and $l$ (cfr \cite{Haw} or \cite{Ram}),
\[
\log Z_{\mathcal{B}_D}(T,q,l)
=\frac{l^D}{\pi^\frac{D+1}{2}}\Gamma\left(\frac{D+1}{2}\right)\zeta_R(D+1)T^D
+O(T^{D-1}),
\]
\[
\log Z_{\mathcal{T}_D}(T,q,l)
=\frac{(2l)^D}{\pi^\frac{D+1}{2}}
\Gamma\left(\frac{D+1}{2}\right)\zeta_R(D+1)T^D
+O(T^{D-1}).
\]

\subsection{Critical volume} \label{s35} As anticipate in the introduction, we
will analyze in this section the dependence on the volume of the
pressure of the radiation at finite temperature. The analysis is
performed for the two models described in section 4, that only
differ for the boundary conditions: periodic or of Dirichlet type.
For simplicity, just consider the zero mass case. By definition,
\[
\hspace{-2.2cm} P(T,V)=\frac{\partial }{\partial V}T \log
Z(T,q,l)= \frac{1}{2}\frac{\partial }{\partial V}(T\zeta'(0;2\pi
T, 0,l))-\frac{1}{2} \frac{\partial }{\partial V}(\log \rho \ T
\zeta(0;2\pi T,0,l) ).
\]

Notice that applying corollary \ref{c6} for the Dirichlet boundary
conditions, we always have the vanishing of the zero mass zeta
function at $s=0$, namely $\zeta(0;2\pi T,0,l)=0$ for the box. For
periodic boundary condition, we get a non trivial term involving the
renormalization constant $\rho$. Thus the analysis in the following
holds for the torus only if we assume the renormalization constant to
be volume independent \footnote{As observed in the introduction, we are
not going to analyze here the renormalization aspects of the model.
Beside, notice that even with a renormalization parameter depending on
the volume, the volume effect still exists, but critical volume
depends also on the explicit form of $\rho$. This one, as pointed out
in \cite{ET}, gives the connection between the model and the physical
reality.}. With this assumption, we get at each fixed temperature
\[
P_{\mathcal{T}_D}(T,V)=\frac{2\pi}{D} V^{-\frac{D+1}{D}}
\left(\sum_{k\in \mathbb{Z}_0^D} \frac{|k|}{\e^\frac{2\pi
|k|}{TV^\frac{1}{D}}-1}-H_D\right),
\]
\[
P_{\mathcal{B}_D}(T,V)=\frac{\pi}{D} V^{-\frac{D+1}{D}}
\left(\sum_{k\in (\mathbb{N}_0)^D} \frac{|k|}{\e^\frac{\pi
|k|}{TV^\frac{1}{D}}-1}-K_D\right),
\]
where the constants are
\[
H_D=-\frac{1}{2}\hat\zeta_D\left(-\frac{1}{2};0\right),\hskip .3in
K_D=-\frac{1}{2}\xi_D\left(-\frac{1}{2};0\right).
\]

We show that for the periodic boundary condition there is, for all
$D$, a value $V_0$ of the volume where the pressure changes sign,
being attractive when $V<V_0$. We also show that the same happens
for the Dirichlet boundary condition at the physical dimension
$D=3$. For the torus, consider the function
\[
g_D(x)=\sum_{k\in \mathbb{Z}_0^D} \frac{|k|}{\e^\frac{ |k|}{x}-1}
=\frac{\sqrt{D}}{\e^\frac{ \sqrt{D}}{x}-1}+g_0(x),
\]
for $x>0$ and the following inequality
\begin{lem}\label{lvc} If $x>ab$  and $y\geq b$ ($x,y,a,b>0$),
then
\[
\e^{\frac{ab}{a+1}y}\left(\e^{\frac{ab}{1+a}x}-1\right)<\left(\e^{xy}-1\right).
\]
\end{lem}
Taking $b=1$ and $a=\frac{1}{x_0}$, we get the following bounds
for $g_D(x)$, when $x<x_0$,
\[
\frac{\sqrt{D}}{\e^\frac{\sqrt{D}}{x}-1}<g_D(x)<
\frac{\sqrt{D}}{\e^\frac{\sqrt{D}}{x}-1}+\frac{1}{\e^{\frac{1}{1+x_0}\frac{1}{x}}-1}
C_D(x_0),
\]
where
$C_D(x_0)=\sum_{k\in\mathbb{Z}^D_{\{0,1\}}}|k|\e^{-\frac{1}{1+x_0}|k|}$,
is a positive constant. It is thus clear that the pressure changes
sign for some value of $V$ (and fixed $T$) if the constant $H_D$
is positive. To show that this is the case, just use the
reflection formula (lemma 5) that for the function
$\hat\zeta_D(s;0)$ takes the simpler form
\[
\pi^{-s}\Gamma(s)\hat\zeta_D(s;0)
=\pi^{s-\frac{D}{2}}\Gamma\left(\frac{D}{2}-s\right)\hat\zeta_D\left(\frac{D}{2}-s;0\right).
\]

Using lemma \ref{lvc} with $b=2$ for the box, we get for the
function
\[
f_D(x)=\sum_{k\in (\mathbb{N}_0)^D} \frac{|k|}{\e^\frac{
|k|}{x}-1} =\frac{\sqrt{D}}{\e^\frac{ \sqrt{D}}{x}-1}+f_0(x),
\]
($x>0$) similar bounds when $x<\frac{x_0}{2}$:
\[
\frac{\sqrt{D}}{\e^\frac{\sqrt{D}}{x}-1}<g_D(x)<
\frac{\sqrt{D}}{\e^\frac{\sqrt{D}}{x}-1}+\frac{1}{\e^{\frac{2}{1+x_0}\frac{1}{x}}-1}
L_D(x_0),
\]
where $ L_D(x_0)=\sum_{k\in(\mathbb{N}_0)^D-\{1\}}
|k|\e^{-\frac{2}{1+x_0}|k|}$, is a positive constant. Furthermore,
using simple bounds for the norm, we get a bound for the constant
$L_D(x_0)$
\[
L_D(x_0)\leq 2^\frac{D-1}{2}
\frac{\e^{-\frac{4D}{1+x_0}}\left(2-\e^{-\frac{2}{1+x_0}}\right)^D}
{\left(1-\e^{-\frac{2}{1+x_0}}\right)^{2D}}.
\]

We can not prove that $K_D$ is positive in general, but we can analyze
explicitly the low dimensional cases. We find that $K_D$ is negative
for $D=2$, but it is positive for $D=1$ and $3$. This indicates that,
up to renormalization, the presence/absence of a boundary does not
affect the Casimir effect in the physical dimension for the model
under study. Eventually, using the above bounds, we can provide bounds
for the solution $x=x_*$ of $f_D(x)=K_D$. For example, if $D=3$, we get
$K_3=0.0050075$, and with $x_0=1.5923$, $C_3(x_0)\leq 0.23433$, and
$0.19684<x_*<0.29613$.

\vskip2truecm
{\bf Acknowledgements}\\
The authors thank the referees for useful suggestions and
bibliographical references.

\end{document}